# Denoising Auto-encoding Priors in Undecimated Wavelet Domain for MR Image Reconstruction

Siyuan Wang, Junjie Lv, Yuanyuan Hu, Dong Liang, *Senior Member, IEEE*,
Minghui Zhang, Qiegen Liu, *Member, IEEE*

*Abstract*—Compressive sensing is an impressive approach for fast MRI. It aims at reconstructing MR image using only a few under-sampled data in k-space, enhancing the efficiency of the data acquisition. In this study, we propose to learn priors based on undecimated wavelet transform and an iterative image reconstruction algorithm. At the stage of prior learning, transformed feature images obtained by undecimated wavelet transform are stacked as an input of denoising autoencoder network (DAE). The highly redundant and multi-scale input enables the correlation of feature images at different channels, which allows a robust network-driven prior. At the iterative reconstruction, the transformed DAE prior is incorporated into the classical iterative procedure by the means of proximal gradient algorithm. Experimental comparisons on different sampling trajectories and ratios validated the great potential of the presented algorithm.

## I. INTRODUCTION

Magnetic resonance imaging (MRI) is an important means of medical imaging. Compared with other radiological methods, MRI has obvious advantages in soft tissue imaging and radiological hazards. Nevertheless, MRI is associated with an inherently slow acquisition speed that is due to data samples not being collected directly in image space but rather in k-space. This slow acquisition could lead to artifacts, due to patients' movement and physiological motion. Compressed sensing MRI (CS-MRI) [1]-[5] enables fast acquisition. i.e., it undersamples k-space and provides an acceleration rate that is inversely proportional to the traversals required. Then, the undersampling measure is reconstructed by nonlinear optimization or iterative algorithms.

The desired solution of the CS-MRI can be achieved by following minimization:

$$\min_u \left\| F_p u - f \right\|^2 + \lambda R(u) \qquad (1)$$

where $\lambda$ is the penalty parameter. This model consists of the data-fidelity term and the regularization term. In the data-consistency component, $F_p$ represents the partially Fourier encoding matrix and $f$ is the raw measurement data in k-space. On the other hand, the regularization term $R(u)$ is the key to reduce imaging artifacts and improve imaging precision. In most early CS-MRI methods, it is often equipped with explicit prior formulation such as the sparsity-enforcing prior. Sparse regularization can be explored in specific sparsifying transform domain or general dictionary-based subspace [6]-[11]. For instance, total variation (TV) in gradient domain has been widely utilized [9], [11]. However, this method introduces staircase artifacts into the reconstruction image. Likely, approaches on the basis of wavelet transform cannot achieve satisfying results, due to the lack of adaptiveness.

Deep learning is widely used in machine learning for various tasks [12]-[23]. Similarly, good results have been achieved in solving CS-MRI problems. For instance, Wang *et al.* [17] learned an end-to-end mapping using convolutional neural networks (CNNs) and applied it to accelerate MRI. These works were followed by novel extension using deep residual learning [19]. For example, Schlemper *et al.* proposed a cascade of network, which simulated the iterative reconstruction of dictionary learning-based methods. Subsequently, they extended the network for dynamic MR reconstructions [21]. Hammernik *et al.* introduced a variational network (VN) for effective reconstruction of complex multi-coil MR data, which inserted the concept into deep learning framework [20]. Inspired by Generative Adversarial Networks (GANs)'s success in image processing community, Mardani *et al.* had employed in reconstructing zero-filling under-sampled MRI, which considered the data consistency during the training process [22]. In particular, Yang *et al.* proposed a new condition-based antagonistic network model (DAGAN) to reconstruct MRI [23]. In DAGAN architecture, an improved learning method was designed to stabilize the generator based on U-network. It provided an end-to-end network to reduce artifacts.

Most of the above-mentioned deep learning methods are conducted in an end-to-end principle, enforcing an implicit prior on reconstruction task. However, these methods lack interpretability and flexibility. In order to inherit the strengths of classical sparsity-enforcing algorithms and the recent supervised learning algorithms, we propose a wavelet transform guided denoising autoencoder prior for CS-MRI, termed WDAEP. Specifically, motivated by some recent works on the wavelet domain with deep learning approaches [24]-[25], our work introduces a novel image prior inheriting the strengths of undecimated wavelet transform and CNN schemes. Particularly, we first learn a denoising autoencoder (DAE) [27] as the prior in undecimated wavelet domain. The autoencoder-based prior learns from a noisy intermediate image to remove the artifacts. Then, at the phase of MRI

This work was supported in part by the National Natural Science Foundation of China under 61871206, 61661031 and project of innovative special funds for graduate students in Jiangxi province (CX2019086).

S. Wang, J. Lv, M. Zhang and Q. Liu are with the Department of Electronic Information Engineering, Nanchang University, Nanchang 330031, China. ({wangsiyuan, lvjunjie}@email.ncu.edu.cn, {zhangminghui, liuqiegen}@ncu.edu.cn).

Y. Hu and D. Liang are with Paul C. Lauterbur Research Center for Biomedical Imaging, Shenzhen Institutes of Advanced Technology, Chinese Academy of Sciences, Shenzhen 518055, China (huyuanyuan98@163.com, dong.liang@siat.ac.cn).

reconstruction, we embed the transformed prior information into the iterative reconstruction process.

To the best of our knowledge, this is the first work that investigates the DAE priors in transform domain. The contributions of this work are summarized as follows:

- Undecimated wavelet transform is used to convert the object from image domain to feature domain, subsequently, a multi-channel and multi-scale tensor is formed as the network input for learning DAE prior, namely WDAEP. By exploiting the redundant and complementary information, the capability of representation prior information is improved.
- In order to reduce the calculation amount of the gradient in WDAEP, a more efficient network is introduced to replace the original encoder network by mathematical equivalence. The new network maps noisy tensor object to purely Gaussian noise.
- The mathematical model WDAEPRec is tackled by proximal gradient descent, followed by integrating the learned WDAEP into classical CS-MRI. At iterative reconstruction stage, the intermediate result is firstly mapped to higher-dimensional space by wavelet transform and then priors provide promising estimation.

The remainder of the paper is organized as follows. Section II provides a brief description of preliminary work with regard to denoising autoencoder prior and wavelet transform. Section III presents the WDAEP model and the corresponding iterative solver. Extensive experimental comparisons among the proposed WDAEPRec and state-of-the-art methods are conducted in Section IV. Finally, concluding remarks and directions for future research are given in Section V.

## II. PRELIMINARIES

### A. Denoising Autoencoder Prior

Denoising autoencoder prior (DAEP) [26] leverages a neural autoencoder to define the image prior [27]-[31]. By denoting DAE as $A_{\sigma_\eta}$ and $u$ as the input image, then the output $A_{\sigma_\eta}(u)$ is trained by adding artificial Gaussian noise $\eta$ with an expected quadratic loss:

$$L_{DAE} = E_{\eta,u}[\|A_{\sigma_\eta}(u+\eta)-u\|^2] \quad (2)$$

where the expectation $E_{\eta,u}[\circ]$ is conducted over all images $u$ and noise $\eta$ with standard deviation $\sigma_\eta$.

According to [27], the network output $A_{\sigma_\eta}(u)$ is related to the true data density $q(u)$ as follows:

$$A_{\sigma_\eta}(u) = u - \frac{\int g_{\sigma_\eta}(\eta) q(u-\eta)\eta d\eta}{\int g_{\sigma_\eta}(\eta) q(u-\eta) d\eta} \quad (3)$$

where $g_{\sigma_\eta}(\eta)$ represents a local Gaussian kernel with standard deviation $\sigma_\eta$. As indicated, the output of an optimal DAE $A_{\sigma_\eta}(u)$ is a local mean of the true data density and then its error is a mean shift vector [31].

By means of the Gaussian derivative definition in $g_{\sigma_\eta}(\eta)$, it yields

$$A_{\sigma_\eta}(u) - u = \sigma_\eta^2 \nabla \log[g_{\sigma_\eta} * q](u) \quad (4)$$

i.e., the autoencoder error $A_{\sigma_\eta}(u) - u$ is proportional to the gradient of the log likelihood of the smoothed density with the image likelihood $Prior(u) = \log \int g_{\sigma_\eta}(u) q(u+\eta) d\eta$. Hence, DAEP utilizes the migratory characteristic of prior information $R(u)$ and uses the magnitude of this mean shift vector as the negative log likelihood of the image prior.

The autoencoder error vanishes at stationary points, including local extrema, of the true density smoothed by the Gaussian kernel. Therefore, the squared magnitude of the autoencoder error is naturally utilized as a regularization prior, i.e., $\|A_{\sigma_\eta}(u) - u\|^2 = \|\sigma_\eta^2 \nabla \log[g_{\sigma_\eta} * q](u)\|^2$. In this work, we propose to exploit the wavelet transform in the basic DAE to form a more efficient prior.

### B. Wavelet Transform

Wavelet transform (WT) has a long history in engineering, such as image coding and denoising [32]-[37]. It packs most of the signal energy into a few significant coefficients and decorrelates random processes into nearly independent coefficients. The WT operator is able to improve the denoising efficiency and preserve or even enhance the edge features.

After the first wavelet soft-thresholding approach proposed by Donoho *et al.* [38], many wavelet-based regularizers have been developed [39]-[46]. These methods can be roughly divided into two parts: the down-sampling wavelet transform (DWT) [45] and the undecimated wavelet transform (UWT) [39], [43]. DWT is an implementation of the wavelet transform using a discrete set of wavelet scales and translations. The sub-bands in DWT are generated by one low-pass filter and three high-pass filters (i.e., horizontal, vertical, and diagonal). The transform properties are as the same as DWT, except for the decimation. The size of each sub-band in UWT is four times than that of DWT. Moreover, the sub-bands decomposed under UWT are redundant. This transform decomposes the input into one global average plus difference signals, achieving better results in image representation and reconstruction. Therefore, we select UWT to encode the input image in this work. To better understand the difference between DWT and UWT, the decomposition procedures for them are depicted in Fig. 1.

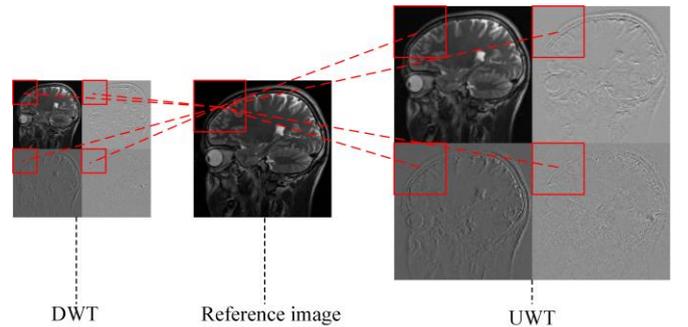

Fig. 1. The 2D-WT decomposition. Left: The 2D-DWT decomposition, Middle: Reference image, Right: The 2D-UWT decomposition. Haar wavelet is used in both WTs.

## III. PROPOSED WDAEP MODEL

In this section, a WDAEP model and the iterative reconstruction WDAEPRec algorithm are presented. First, we develop a WDAEP model from the naive DAEP, which is trained on the feature images generated by undecimated wavelet transform. Second, alternative iterative strategy and proximal gradient method are jointly adopted to address the

WDAEP-based CS-MRI model. At the iterative stage, the intermediate result is firstly mapped to higher-dimensional space by forward wavelet transform. The priors in transform domain are used to provide promising estimation. Then the inverse wavelet transform returns it to the data-fidelity iterative scheme.

A. *Wavelet-guided DAEP (WDAEP)*

Wavelets are orthogonal basis functions that decompose data into different spatio-frequency components. Due to the delineating capability, wavelets have a better discrimination between the noise and the signal. Therefore, wavelet methods have a strong impact on the field of image processing, especially in image coding and image denoising.

In terms of formula derivation, wavelets are mathematical functions, which are generated from one basic function. If the basic wavelet is denoted by $\psi(t)$, other wavelets $\psi_{a,b}(t)$ can be represented as:

$$\psi_{a,b}(t) = 1/\sqrt{|a|} * \psi((t-b)/a) \tag{5}$$

where $a$ and $b$ are two real numbers which represent the parameters for dilations and translations respectively in the time axis. The scale parameter $a$ and the shift parameter $b$ are given by $2^{-j}$ and $k2^{-j}$, where $j,k \in Z$.

Therefore, the family of wavelet functions is represented as:

$$\psi_{j,k}(t) = 2^{j/2}\psi(2^j t - k) \tag{6}$$

Analyzing wavelet transform in multi-scale view, the decomposition of a discrete time signal $x[n]$ is given as:

$$x[n] = \sum_k W_\varphi(j_0, k)\varphi_{j_0,k}(n) + \sum_{j=j_0}^{\infty} \sum_k W_\psi(j,k)\psi_{j,k}(n) \tag{7}$$

where $\varphi_{j_0,k}$ and $\psi_{j,k}$ are the scaling (low-pass) functions and wavelet (high-pass) functions respectively. The scaling and the wavelet coefficients are given as:

$$W_\varphi(j_0, k) = \sum_n x[n] 2^{\frac{j_0}{2}} \varphi(2^{j_0}n - k) \tag{8}$$

$$W_\psi(j, k) = \sum_n x[n] 2^{\frac{j}{2}} \psi(2^j n - k) \tag{9}$$

where $j_0$ is the starting scale always being zero. The DWT is always defined as two-channel sub-band decomposition. The forward and inverse relations are given as:

$$W_\varphi(j,k) = h_\varphi(-n) * W_\varphi(j+1,n)\big|_{n=2k; k\geq 0}$$
$$W_\psi(j,k) = h_\psi(-n) * W_\varphi(j+1,n)\big|_{n=2k; k\geq 0} \tag{10}$$

$$W_\varphi(j+1,k) = \tilde{h}_\varphi^{(k)} * W_\varphi^{up}(j,k) + \tilde{h}_\psi^{(k)} * W_\psi^{up}(j,k)\big|_{k\geq 0} \tag{11}$$

In UWT, the solutions are undecimated and two new sequences have the same length as the original sequence. The relationship between scales is given by:

$$W_\varphi(j,k) = h_\varphi(-n) * W_\varphi(j+1,n)\big|_{n=k; k\geq 0}$$
$$W_\psi(j,k) = h_\psi(-n) * W_\varphi(j+1,n)\big|_{n=k; k\geq 0} \tag{12}$$

$$W_\varphi(j+1,k) = \tilde{h}_\varphi^{(k)} * W_\varphi^{up}(j,k) + \tilde{h}_\psi^{(k)} * W_\psi^{up}(j,k) \tag{13}$$

In this work, by incorporating the regularization in transformed domain as a whole into the data-consistency term, we propose a wavelet transform-guided DAEP (WDAEP):

$$\min_u \|F_p u - f\|^2 + \lambda \|A_{\sigma_\eta}(\Phi(u)) - \Phi(u)\|^2 \tag{14}$$

where $\Phi(u) = W(u)$, $W$ stands for the wavelet transform. The superiority of Eq. (14) can be derived from the following equality:

$$\|A_{\sigma_\eta}(\Phi(u)) - \Phi(u)\|^2 = \|A_{\sigma_\eta}(W(u)) - W(u)\|^2$$
$$= \|\sigma_\eta^2 \nabla \log[g_{\sigma_\eta} * q](W(u))\|^2 \tag{15}$$

Comparing the derivation in Eq. (15) and Eq. (4), it can be observed that the proposed method exploits a higher-dimensional probability.

After the model is formed, an emergency issue is to minimize it via some gradient descent techniques. Nevertheless, directly calculating the gradient of Eq. (14) involves too complicated operators. Alternatively, by introducing a new network $D_{\sigma_\eta}(\Phi(u)) = \Phi(u) + \eta - A_{\sigma_\eta}(\Phi(u))$, an equivalence of Eq. (14) is presented as follows:

$$\min_u \|F_p u - f\|^2 + \lambda \|D_{\sigma_\eta}(\Phi(u)) - \eta\|^2 \tag{16}$$

As can be seen in the regularization term of Eq. (16), its gradient only involves one variable, which largely reduces calculation amount.

Correspondingly, at the prior learning stage, the target of the output in $D_{\sigma_\eta}(\Phi(u))$ becomes to be Gaussian noise as close as possible via the popular L2 loss function, i.e.,

$$L_{WDAE} = E_{\eta,u}[\|D_{\sigma_\eta}(\Phi(u)) - \eta\|^2] \tag{17}$$

The detailed description of the network $D_{\sigma_\eta}(\Phi(u))$ will be given in next subsection.

B. *Network $D_{\sigma_\eta}(\Phi(u))$*

In the process of training $D_{\sigma_\eta}(\Phi(u))$, the wavelet transform UWT servers as a "bridge" between image domain and wavelet domain. After applying UWT to original image, the wavelet coefficients from 4-subbands at the same spatial location forms a tensor as the network input. Such an operation naturally incorporates the dependencies of wavelet coefficients to improve the representation ability.

As shown in Fig. 2, the network architecture of $D_{\sigma_\eta}(\Phi(u))$ contains the following main characteristics: an UWT serves as the encoder of the whole network. "Conv + ReLU" block, "Conv + BN + ReLU" block, and the residual "Conv + BN + ReLU" block, which adds a sum layer from the first "Conv" to the last "BN". Note that the abbreviation "Conv" represents a convolutional layer, "BN" denotes the batch normalization, and "ReLU" represents the rectified linear unit. For the network input, we concatenate the real and imaginary feature maps of the complex image in wavelet domain, and an eight-channel tensor is finally considered by adding random Gaussian noise. Specifically, assuming the size of the complex image is $C^{m \times n}$. After converting it into real and imaginary components, the data size becomes to $C^{m \times n \times 2}$. Through UWT, the size of the input tensor in WDAE network is $R^{m \times n \times 8}$.

Except for the last layer, the kernel number of each convolutional layer is set to be 320. The kernel number of the last layer is set to be 8. The kernel size of each convolutional layer is $3 \times 3$. In addition, zero padding is adopted to keep all feature maps having the same size in all layers.

One of the most important parameters in WDAEP is the noise level of the added Gaussian noise at the network input.

As well known, the network in DAE [27]-[30] is trained to reconstruct each data point from a destroyed version. The noise-adding process is chosen by users and has much influence on the final representation. Glorot *et al.* noticed that using either too low or too high level of noise in network training will degrade the representation accuracy [29]. In most applications of representation learning, excellent representations should contain features of various sizes to learn multi-scale features. Geras and Sutton introduced scheduled denoising autoencoders [30], which are based on the intuition that the features at different scales can be learned by training the same network at multiple noise levels. Although this idea is promising, training a series of network is highly time-consuming. In this work, by adding artificial noise with the same level to the feature maps with multi-scale and multi-resolution information, we achieve the same goal of learning a DAE network that exploiting multi-scale information.

As discussed, the network $D_{\sigma_\eta}(\Phi(u))$ removes the latent clean image from noisy observation and the output approaches to Gaussian noise. In Fig. 3, the convolution filters from five different layers are visualized. At the first layer, it can be observed that many features appear at different levels of granularity, such as coarse-grained and fine-grained. As the layer increases, more and more filters containing noise-like patterns occur at the middle layers. Finally, at the 20-th layer (i.e., the last layer), it is noticed that many edge-preserving patterns become to be orientation-free and random.

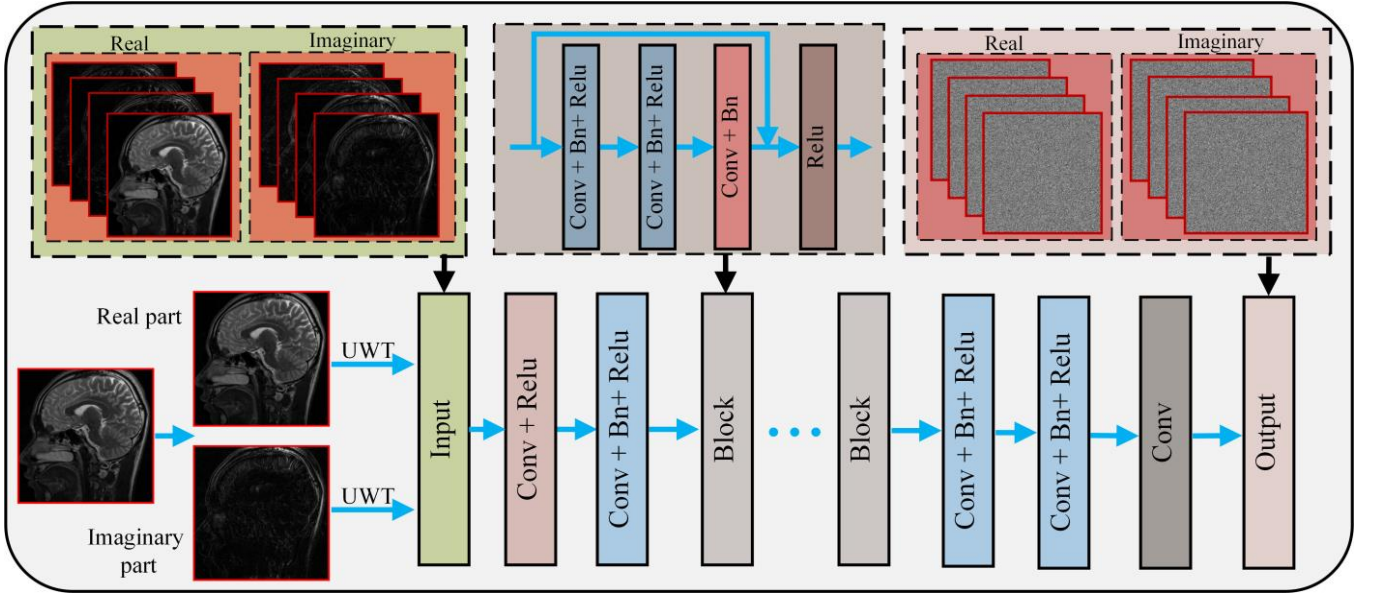

Fig. 2. The architecture of the proposed network $D_{\sigma_\eta}(\Phi(u))$. The "Block" (dark green) is used to describe heavy network units with residual structure. Detailed structure of "Block" is specified in the dotted box. Five residual blocks are used in our network.

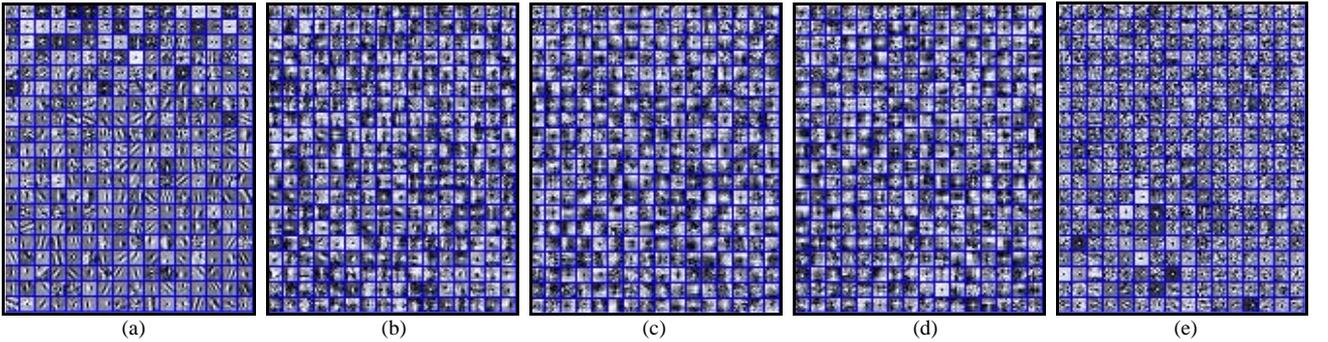

(a)　　　　　(b)　　　　　(c)　　　　　(d)　　　　　(e)

Fig. 3. Visualization of convolution filters in WDAEP at noise level $\sigma_\eta = 25$. From left to right: the convolution kernels generated by WDAEP at 1-st, 3-rd, 8-th, 14-th, and 20-th convolutional layers, respectively.

## C. WDAEPRec: Iterative Solver for WDAEP

The general mathematical WDAEP model for CS-MRI reconstruction, termed WDAEPRec, can be derived as follows:

$$\min_u \|F_p u - f\|^2 + \lambda \|D_{\sigma_\eta}(\Phi(u)) - \eta\|^2 \quad (18)$$

where the second term consists of the network-driven prior information. Due to the nonlinearity of the model, we apply the proximal gradient method [37] to tackle it. The model is approximated by a standard least square (LS) minimization:

$$\min_u \|F_p u - f\|^2 + \frac{\lambda}{\beta} \|u - (u^k - \beta \nabla G(u^k))\|^2 \quad (19)$$

where $G(u) = \|D_{\sigma_\eta}(\Phi(u)) - \eta\|^2$ and $\nabla G(u) = \Phi^T \{\nabla D_{\sigma_\eta}^T (\Phi(u)) [D_{\sigma_\eta}(\Phi(u)) - \eta]\}$.

The function $G(u)$ is $1/\beta$-Lipschitz smooth, i.e., $\|\nabla G(u') - \nabla G(u'')\|_2 \leq \|u' - u''\|_2 / \beta$. $k$ denotes the index number of iterations. Here, we empirically set $\beta = 1$ and it achieves outstanding performance in our experiments.

Given $\beta=1$, Eq. (19) can be solved as follows:

$$u^{k+1} = \frac{F_p^T f + \lambda[u^k - \Phi^T\{\nabla D_{\sigma_\eta}^T(\Phi(u^k))[D_{\sigma_\eta}(\Phi(u^k))-\eta]\}]}{(F_p^T F_p + \lambda)} \quad (20)$$

The key to solve $u^{k+1}$ is to update $u^k - \Phi^T\{\nabla D_{\sigma_\eta}^T(\Phi(u^k))[D_{\sigma_\eta}(\Phi(u^k))-\eta]\}$. Noted that the parameters in $D_{\sigma_\eta}(\circ)$ have already learned in the network training stage. More specifically, the $D_{\sigma_\eta}(\Phi(u^k))$ is the forward output of the network and $\eta$ is the artificial noise. $\nabla D_{\sigma_\eta}^T(\circ)$ is the derivative which can be solved by the backward of the network. $\nabla D_{\sigma_\eta}^T(\Phi(u^k))[D_{\sigma_\eta}(\Phi(u^k))-\eta]$ is the network backward output with the input $D_{\sigma_\eta}(\Phi(u^k))-\eta$. In brief, the mathematical model is tackled by the proximal gradient and alternative optimization.

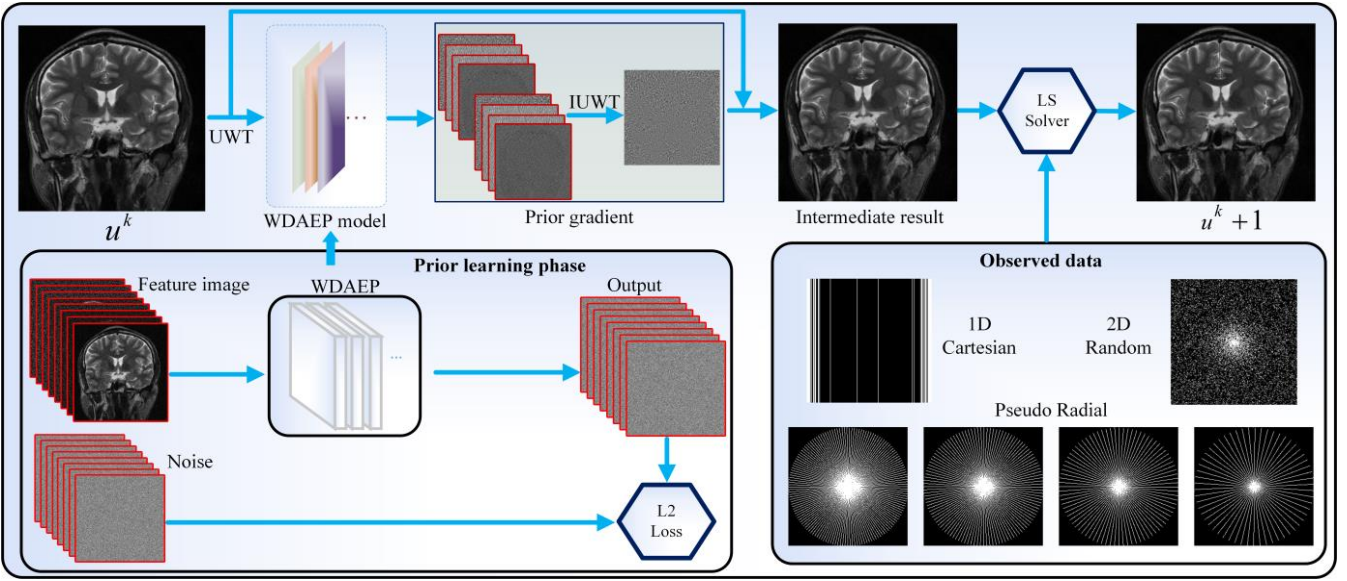

Fig. 4. Schematic flowchart of the proposed WDAEPRec algorithm. At the training stage, we train the WDAEP model to get the prior. Then, the learned prior is used at the iterative reconstruction phase. Various sampling masks are used in our experiment. Top line: Cartesian sampling and variable density 2D Random sampling with the same acceleration factor $R=6.7$; Bottom line: Pseudo Radial sampling with the different acceleration factors $R=4, 5, 6.7, 10$.

Fig. 4 visualizes the schematic flowchart of the proposed WDAEPRec algorithm. We update the solution $u^k$ by prior estimation $\nabla D_{\sigma_\eta}^T(\Phi(u^k))$, $D_{\sigma_\eta}(\Phi(u^k))-\eta$ and the LS solver until solution $u$ converges. The overall training phase and testing phase for employing WDAEP in WDAEPRec are as follows:

| |
|---|
| **Training stage** |
| **Training images**: Image dataset $\{\Phi(u)\|\Phi(u)=W(u)\}$ |
| **Outputs**: Trained network $D_{\sigma_\eta}(\circ)$ with noisy level $\sigma_\eta$ |
| **Testing stage** |
| **Initialization:** $u^0 = F_p^T f$; $K$ |
| **For** $k=1, 2, \cdots, K$ **do** |
|     Update the variable in wavelet: $\Phi(u^k)=W(u^k)$ |
|     Calculate the prior gradient components: |
|     $D_{\sigma_\eta}(\Phi(u^k))$, $\nabla D_{\sigma_\eta}^T(\Phi(u^k))[D_{\sigma_{\eta i}}(\Phi(u^k))-\eta]$ |
|     Update the LS solution via Eq. (19) |
| **End** |

### D. Relation to EDAEP and VAE

In ref. [47], Liu *et al.* proposed an enhanced DAE prior, dubbed EDAEP. EDAEP improves the reconstruction quality by channel-copy strategy. i.e., in order to reconstruct a single-channel image, it firstly learns a higher-dimensional EDAE prior with three-channel images as input. Then at the reconstruction procedure of applying EDAEP to a single-channel intermediate image, it uses channel-copy technique to map EDAEP space and then averages the three-channel outputs to get a single-channel output. The forward and inverse transformations are as follows:

$$[1\ 1\ 1]^T \times u = [u\ u\ u]^T \equiv U \quad (21)$$
$$[1\ 1\ 1] \times U/3 = u \quad (22)$$

where $u$ is the single-channel image and $U$ is the three-channel artificial image. Furthermore, it applies two-sigma rule to improve the prior robustness. By contrast, WDAE can extract multiscale feature images by UWT and only use single-sigma scheme. i.e., WDAEP applies UWT ($W(u)=U$) to get high-dimensional priors and inverse UWT ($W^T(U)=u$) to get the single-channel output. In a word, both EDAEP and WDAEP adopt higher-dimensional prior in the single-channel image reconstruction, while using UWT is more efficient than the channel-copy strategy.

The WDAEP is also close to variational autoencoder (VAE) [48, 49]. In general, DAEP, WDAEP and VAE fall into the category of unsupervised learning. The naïve DAE adds noise at the level of the input image, while the noise in the proposed WDAE is added at the level of transformed feature images. On the other hand, a VAE is composed by an encoder and a decoder network. Additionally, the noise in VAE is added following the encoding layer [50]. From the viewpoint of network architecture, a VAE can be viewed as a denoising compressive autoencoder. The "compressive" means that the middle layers have lower capacity than the outer layers. On the contrary, the WDAE can be regarded as a denoising expansive autoencoder, where the feature images produced by wavelet transform have higher capacity than the original image.

## IV. Experimental Results

In this section, the performance of WDAEPRec is demonstrated at different sampling schemes. In prior learning stage, we use 500 images with size of $256 \times 256$ and set the training patch size as $40 \times 40$. The size of each batch is $40 \times 40 \times 8 \times 128$. The images for training are normalized to a maximum magnitude of 1. We allow 20 epochs in the training procedure that take nearly 100 hours. We evaluate the performance of the proposed method using a variety of sampling schemes with different acceleration rates ($R$) on 31 2D complex-valued MRI data with the size of $256 \times 256$. Fig. 5 shows three fully-sampled representative images from the 31 images. The wavelet transform discrete Meyer (dmey) is used in the experiment.

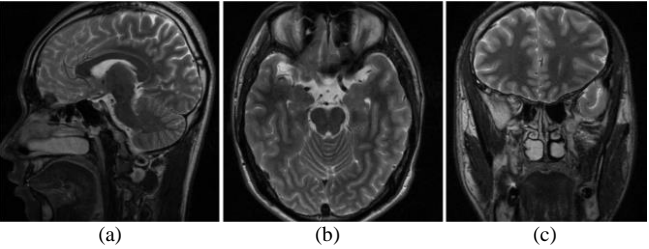

Fig. 5. The representative testing images. (a) Test 1; (b) Test 2; (c) Test 3.

The experiments are implemented in MATLAB (2016a) with the MatConvNet package, an Intel i7-6900K CPU and a GeForce Titan XP GPU. For more in-depth study and research, the source code of WDAEPRec can be found at: https://github.com/yqx7150/WDAEPRec.

To evaluate the quality of the reconstructed image, the popular PSNR (Peak Signal to Noise Ratio, dB), the powerful perceptual quality metrics SSIM (Structural Similarity) [51] and HFEN (High Frequency Error Norm) [52] are calculated. Denoting $u$ and $\hat{u}$ to be the reconstructed image and the ground truth, the PSNR is defined as:

$$PSNR(u,\hat{u}) = 20\log_{10} \text{Max}(\hat{u}) / \|u - \hat{u}\|_2 \quad (23)$$

The SSIM is defined as:

$$SSIM(u,\hat{u}) = \frac{(2\mu_u\mu_{\hat{u}} + c_1)(2\sigma_{u\hat{u}} + c_2)}{(\mu_u^2 + \mu_{\hat{u}}^2 + c_1)(\sigma_u^2 + \sigma_{\hat{u}}^2 + c_2)} \quad (24)$$

Furthermore, HFEN value is to examine the quality of reconstruction of edges and fine features. It is calculated as the ratio between two norms:

$$HFEN = \|LoG(u) - LoG(\hat{u})\|_F^2 / \|LoG(\hat{u})\|_F^2 \quad (25)$$

where the edges are captured via a rotationally symmetric LoG (Laplacian of Gaussian) filter. The size of filter kernel is $15 \times 15$ pixels with a standard deviation of 1.5 pixels.

### A. Comparisons on Different Sampling Rates

The comparisons between WDAEPRec and several state-of-the-art methods are conducted under different sampling rates. The PSNR, SSIM and HFEN values of the obtained results are presented in the left region of Table I. As can be seen, the proposed WDAEPRec method yields the highest values in most majority of the sampling rates. For example, under the accelerate factor $R$=4 and $R$=5, the highest PSNR values achieved by all the compared methods are 34.49 dB and 33.49 dB, which is obviously lower than the values of 35.28 dB and 34.22 dB obtained by WDAEPRec. With regard to the two dictionary-based methods, FDLCP [53] gains 1.63 dB, 1.76 dB, and 1.95 dB improvement than DLMRI [52] at $R$=4, 5, 10, respectively. Besides, the DAE prior incorporated EDAEPRec method gets the best valves, whose PSNR is 0.16 dB higher than WDAEPRec at $R$=10. In a word, WDAEPRec outperforms all the competing approaches expect at extremely high acceleration rates.

Fig. 6 shows the performance of the seven methods under pseudo radial sampling of k-space with $R$=5. Moreover, an enlarged area is presented to reveal the details preserved by each algorithm. The dictionary learning based optimization, such as DLMRI and FDLCP, are insufficient to provide smaller structural features, due to the limitation of a relatively small amount of learnable filters. The reconstructions of these two approaches in Fig. 6 are over-smoothed compared to other algorithms. The patch-based method PANO [54] does not lead to a sufficient sparse representation, and thus lacks sharp edges in the reconstruction. The reconstruction result of EDAEPRec is seen to be better than the grouped low-rank based method NLR-CS [55] and the supervised end-to-end method DC-CNN [56]. EDAEP trains the network using multi-channel strategy and two-sigma in the iteration, which makes the network more robust. Compared to EDAEPRec, our method is almost devoid of aliasing artifacts and successfully preserves the most details. The reconstruction result is more realistic and closer to the original image.

### B. Comparisons on Different Sampling Patterns

In this experiment, we discuss the performance of various algorithms under different sampling patterns (variable density 2D random sampling, pseudo radial sampling and Cartesian sampling). The quality metrics listed in the right region of Table I imply that WDAEPRec achieves the lowest HFEN and highest PSNR and SSIM among all the methods. Specifically, for the three sampling patterns, the average PSNR values achieved by WDAEPRec are 0.56 dB, 0.68 dB, and 0.52 dB higher than the second-best algorithm EDAEPRec. Additionally, under the pseudo radial sampling, WDAEPRec outperforms DC-CNN by 2.11 dB in terms of PSNR.

Variable density 2D random and Cartesian sampling are employed on Test 2 and Test 3 in Figs. 7 and 8, respectively. For the dictionary-based methods, the fast dictionary learning method on classified patches method FDLCP is seen to outperform the DLMRI method on both sampling patterns. FDLCP employs the similarity and the geometrical directions of patches which contains more details than DLMRI. The reconstructions of the PANO and DC-CNN are lack of textures to some extent. For FDLCP and EDAEPRec, the FDLCP method suffers from significant loss of structures, and EDAEPRec generates results that are much closer to ground-truths while edges are almost preserved. Overall, WDAEPRec outperforms EDAEPRec in terms of edge preservation and aliasing artifacts removal (i.e., incoherent aliasing artifacts for variable density 2D random sampling, and streaking artifacts for Cartesian sampling). From the zoom-in reconstructing results of the under-sampling observation, it can be concluded that WDAEPRec outperforms the other methods in reconstructing the fine textures. Particularly, only our algorithm well reconstructs the vessels in the brain white matter, as indicated in the enlarged area.

TABLE I
AVERAGE PSNR, SSIM AND HFEN VALUES OF RECONSTRUCTING 31 TEST IMAGES BY DIFFERENT ALGORITHMS AT RADIAL SAMPLING TRAJECTORIES AND DIFFERENT SAMPLING TRAJECTORIES WITH THE SAME PERCENTAGE.

| | $R$=4, Radial | $R$=5, Radial | $R$=10, Radial | $R$=6.7, 2D Random | $R$=6.7, Radial | $R$=6.7, 1D Cartesian |
|---|---|---|---|---|---|---|
| **DLMRI** | 32.41/0.8866/0.84 | 31.21/0.8602/1.10 | 27.39/0.7444/2.18 | 27.63/0.7518/2.02 | 29.36/0.8103/1.58 | 26.50/0.7390/2.51 |
| **PANO** | 33.65/0.8995/0.73 | 32.44/0.8777/0.96 | 28.58/0.7805/1.90 | 29.12/0.7964/1.77 | 30.60/0.8372/1.37 | 27.51/0.7683/2.28 |
| **FDLCP** | 34.04/0.8980/0.62 | 32.97/0.8770/0.80 | 29.34/0.7856/1.60 | 30.14/0.8004/1.44 | 31.31/0.8391/1.13 | 27.91/0.7776/2.15 |
| **NLR-CS** | 34.35/0.8938/0.61 | 33.32/0.8812/0.79 | 29.51/0.7845/1.65 | 30.34/0.8087/1.46 | 31.35/0.8494/1.17 | 28.23/0.7798/2.03 |
| **DC-CNN** | 34.07/0.8992/0.69 | 32.68/0.8791/0.95 | 28.39/0.7710/1.93 | 28.78/0.7873/1.83 | 30.57/0.8348/1.38 | 27.05/0.7506/2.44 |
| **EDAEPRec** | 34.49/0.9151/0.64 | 33.49/0.8990/0.79 | **30.30/0.8319/1.40** | 30.68/0.8433/1.31 | 32.00/0.8716/1.05 | 28.85/0.8041/1.81 |
| **WDAEPRec** | **35.28/0.9222/0.51** | **34.22/0.9053/0.66** | 30.16/0.8224/1.47 | **31.24/0.8501/1.23** | **32.68/0.8802/0.90** | **29.37/0.8286/1.67** |

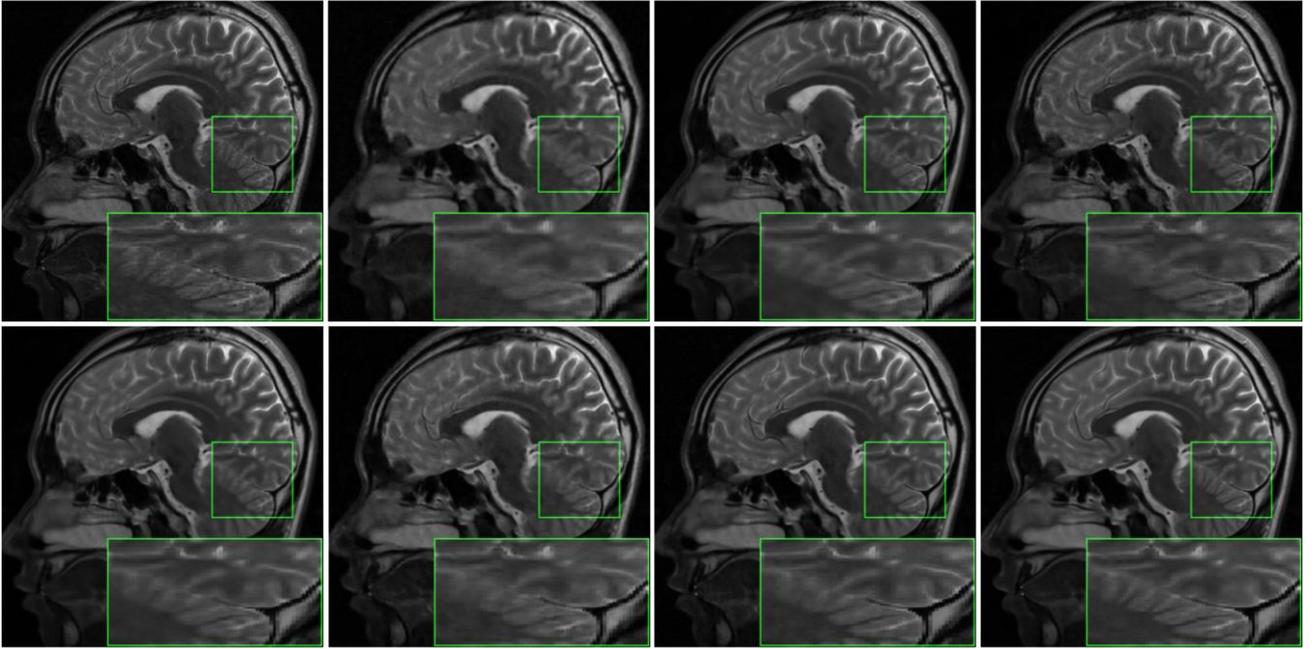

Fig. 6. Visual comparisons under pseudo radial sampling at acceleration factor $R$=5. Top line: reference image, reconstruction results using DLMRI, PANO and FDLCP; Bottom line: reconstruction results using NLR-CS, DC-CNN, EDAEPRec and WDAEPRec.

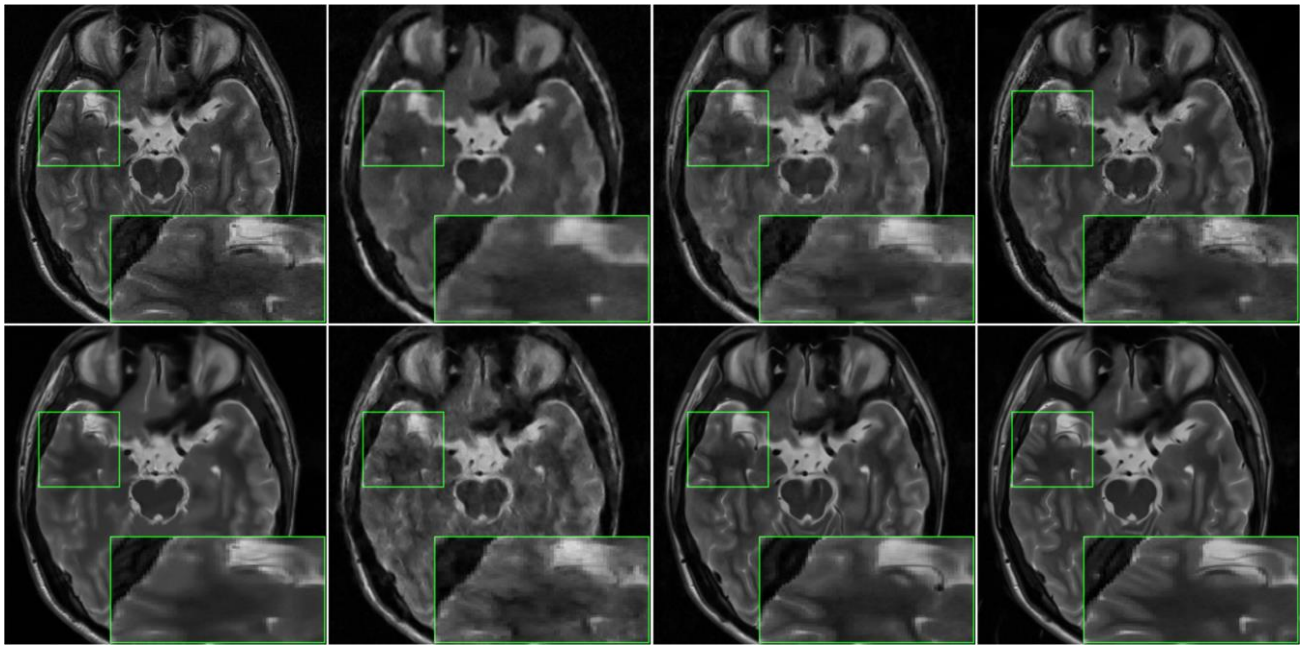

Fig. 7. Visual comparisons under 2D Random sampling at acceleration factor $R$=6.7. Top line: reference image, reconstructed images using DLMRI, PANO and FDLCP; Bottom line: reconstructed images using NLR-CS, DC-CNN, EDAEPRec and WDAEPRec.

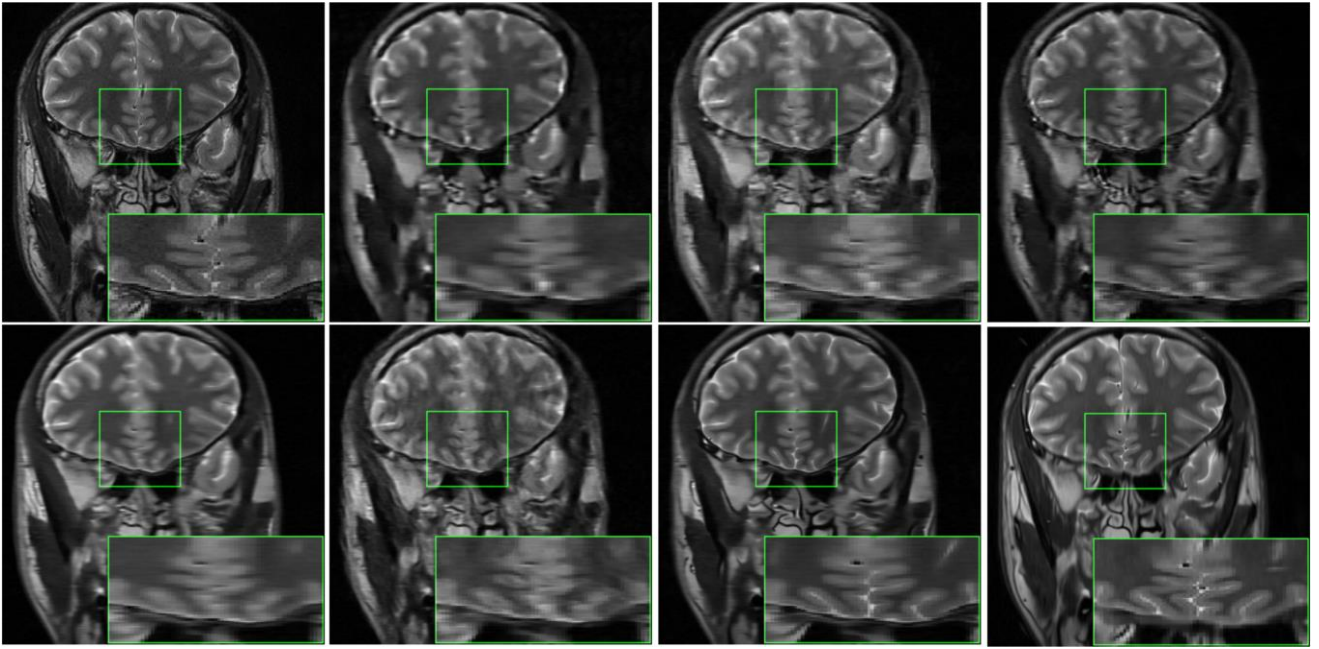

Fig. 8. Performance comparison of the same seven methods with 1D Cartesian sampling ($R$=6.7). Top line: original image, reconstruction results using DLMRI, PANO and FDLCP; Bottom line: reconstruction results using NLR-CS, DC-CNN, EDAEPRec and WDAEPRec.

## C. Variants of Network

In previous experiments, it is empirically shown that reconstructions from undecimated wavelet derived prior can benefit from redundant and complementary information and produce promising results. In this subsection, we investigate the impact of some network parameters during the training procedure: the chosen of wavelet transform, filter number at the layer, residual block number in the network, and the noisy level of the added noise as the network input.

We select different wavelet transform Haar, db4, coif2, sym4 and dmey to analyze the influences of different wavelets on prior information. The results of five undecimated wavelets on Test 1, Test 2 and Test 3 are listed in Table II. As can be seen, Haar and dmey wavelets are the best wavelet candidate bases for our prior reconstruction. On the other hand, db4 and coif2 wavelets derived priors work the worst. As well known, wavelets possess different characteristics that could guide the selection: Compact support, symmetry, regularity, and time-frequency localization. These four factors may contribute to the reconstruction performance. It is worth noting that the superiority of dmey wavelet has been exhibited in many applications [57]-[58]. An example of the reconstructions by WDAEPRec equipped with the five different wavelet transforms is shown in Fig. 9. It is obvious that the reconstructions by Haar, sym4 and dmey wavelets exhibit better detail-preservation than db4 and coif2 wavelets. Particularly, db4 locates the worst position. A more visual illustration of convolution filters learned in WDAEP by db4 wavelet is depicted in Fig. 10. It is noticed that the network learns a lot of fine-grained details, while few global and coarse-grained features. As a result, the network only prefers to small-scale structural information, and leads to fluctuant reconstructions in Fig. 9.

In the second test, the network sensitivity to filter number and block number is examined. In general, the performance would improve if we increase the network width and length. However, more computational time will be cost in the training process if the network becomes bigger. In Table III and IV, we examine the network with filter number of 80, 160, 320, 400 and block number of 1, 2, 5, 7. As the filter number increases from 80 to 320, the reconstruction PSNR gains over 0.86 dB. At the meanwhile, as the residual block increases from 1 to 5, the PSNR gains over 0.64 dB. On the other hand, it is also clear that the improvement becomes very smaller as more filters/blocks are included. In summary, the results indicate that a bigger network could grasp richer structural information, which in turn leads to better results but slower speed. In order to tradeoff between performance and efficiency, we set filter number to be 320 and block number to be 5. Additionally, the PSNRs versus iterations are presented in Fig. 11.

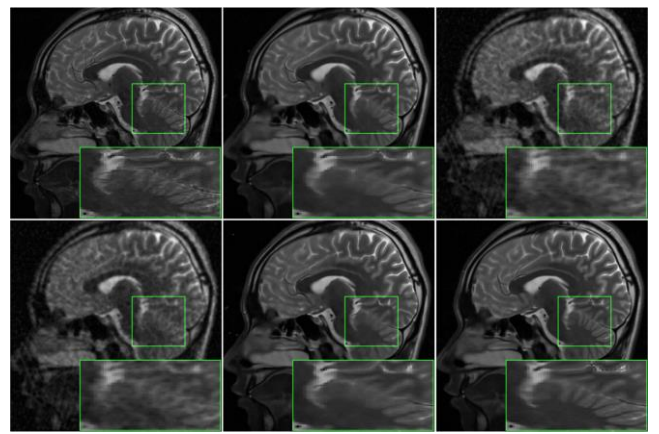

Fig. 9. Reconstructed images using pseudo radial at acceleration factor $R$=6.7 in different wavelet transforms on Test 2. Top line: reference image, Haar and db4 wavelet transforms; Bottom line: coif2, sym4 and dmey wavelet transforms.

TABLE II
AVERAGE RECONSTRUCTION PSNR, SSIM AND HFEN VALUES OF THREE TEST IMAGES RESTORED ON PSEUDO RADIAL SAMPLING WITH $R$=6.7.

| Method | Haar | Db4 | Coif2 | Sym4 | Dmey |
|---|---|---|---|---|---|
| PSNR | 32.58 | 27.57 | 27.54 | 31.90 | **32.59** |
| SSIM | 0.8841 | 0.7356 | 0.7344 | 0.8707 | **0.8842** |
| HFEN | 0.8960 | 2.1805 | 2.1801 | 1.0095 | **0.8912** |

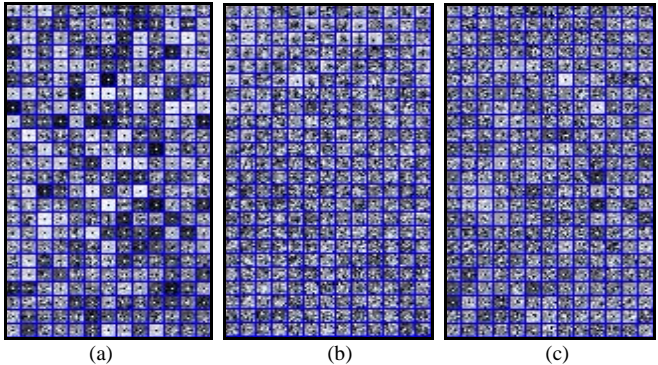

(a)  (b)  (c)

Fig. 10. Visualization of convolution filters with noise level $\sigma_\eta = 25$. The convolution kernels at (a) 1-st; (b) 3-rd; (c) 20-th convolutional layers, respectively. Compared to Fig. 3, the filter features at three layers are few levels of granularity, almost fine-grained features.

TABLE III
AVERAGE RECONSTRUCTION PSNR, SSIM AND HFEN OF THREE TEST IMAGES WITH VARIOUS FILTER NUMBERS ON RADIAL SAMPLING $R$=6.7.

| Filter number | 80 | 160 | 320 | 400 |
|---|---|---|---|---|
| PSNR | 31.73 | 32.43 | 32.59 | **32.61** |
| SSIM | 0.8712 | 0.8818 | 0.8842 | **0.8849** |
| HFEN | 1.0137 | 0.9145 | **0.8912** | 0.8952 |

TABLE IV
AVERAGE RECONSTRUCTION PSNR, SSIM AND HFEN OF THREE TEST IMAGES WITH VARIOUS BLOCK NUMBERS ON RADIAL SAMPLING $R$=6.7.

| Block number | 1 | 2 | 5 | 7 |
|---|---|---|---|---|
| PSNR | 31.95 | 32.37 | 32.59 | **32.61** |
| SSIM | 0.8747 | 0.8792 | 0.8842 | **0.8843** |
| HFEN | 0.9863 | 0.9381 | **0.8912** | 0.8957 |

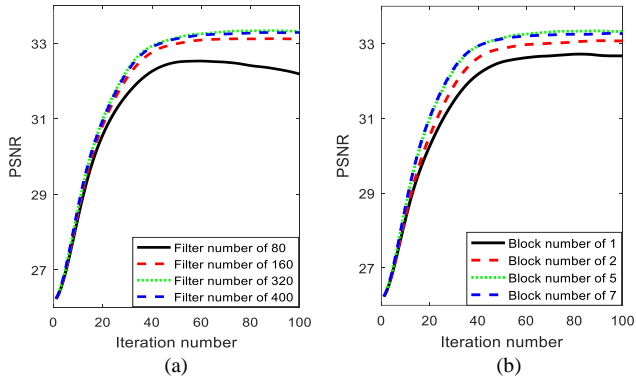

(a)  (b)

Fig. 11. The PSNRs vs. iterations conducted on Test 3. (a) various filter numbers; (b) various block numbers.

Finally, quantitative measures of the proposed algorithm at different noise levels $\sigma_\eta$ = 5, 10, 20, 25, 30, 35 are recorded in Table V. As expected, under extreme noise levels $\sigma_\eta = 5$ and $\sigma_\eta = 35$, WDAEPRec does not guarantee compact and effective representation, subsequently produce degraded reconstruction results. Particularly, the results at $\sigma_\eta = 5$ yield inferior performance compared to other noise levels. On the other hand, setting $\sigma_\eta = 20$ and $\sigma_\eta = 25$ can help achieve the highest PSNR and SSIM performances as well as a satisfactory HFEN performance. In Table VI, we further compare quantitative performance of the abovementioned two noise levels ($\sigma_\eta = 20$ and $\sigma_\eta = 25$) using a variety of sampling schemes on 31 test images. As can be seen, under the accelerate factor $R$=4 and $R$=5, setting $\sigma_\eta = 20$ is able to reconstruct satisfactory results. However, when $R$ increases, setting $\sigma_\eta = 25$ performs better in terms of all the three evaluation metrics: PSNR, SSIM and HFEN. Therefore, in the experiments we choose $\sigma_\eta = 20$ for relatively small accelerate factor values ($R \leq 5$) and $\sigma_\eta = 25$ for larger $R$ values ($R > 5$).

TABLE V
AVERAGE PSNR, SSIM AND HFEN VALUES WITH DIFFERENT $\sigma_\eta$ RESTORED BY WDAEPREC ON PSEUDO RADIAL SAMPLING WITH $R$=6.7.

| $\sigma_\eta$ | PSNR | SSIM | HFEN |
|---|---|---|---|
| 5 | 27.26 | 0.7209 | 2.3196 |
| 10 | 30.36 | 0.8246 | 1.2598 |
| 20 | 32.41 | 0.8762 | 0.8949 |
| 25 | **32.59** | **0.8842** | **0.8912** |
| 30 | 32.40 | 0.8813 | 0.9237 |
| 35 | 32.21 | 0.8789 | 0.9449 |

TABLE VI
AVERAGE PSNR, SSIM AND HFEN VALUES OF RECONSTRUCTING 31 TEST IMAGES.

| Acceleration rates | $\sigma_\eta = 20$ | $\sigma_\eta = 25$ |
|---|---|---|
| $R$=4, Radial | **35.28/0.9222/0.5126** | 34.99/0.9203/0.5363 |
| $R$=5, Radial | **34.22/0.9053/0.6592** | 33.97/0.9041/0.6850 |
| $R$=10, Radial | 29.52/0.8031/1.6542 | **30.16/0.8224/1.4713** |
| $R$=6.7, 2D Random | 30.53/0.8320/1.3634 | **31.24/0.8501/1.2329** |
| $R$=6.7, Radial | **32.76**/0.8792/0.9039 | 32.68/**0.8802/0.9029** |
| $R$=6.7, 1D Cartesian | 28.74/0.8147/1.8163 | **29.37/0.8286/1.6489** |

## V. CONCLUSIONS AND FUTURE WORKS

This work paved a new way to incorporate unsupervised learning derived prior information into the tradition transform. A wavelet transform guided denoising autoencoder WDAE and an induced prior WDAEP were proposed for CS-MRI. Specifically, we made use of the merits of both transform optimization-based and network-based CS methods in a unified framework. The wavelet used in the current work can be naturally extended to other more special-designed transforms such as Curvelet, Contourlet, etc. Furthermore, extending the proposed prior to other imaging modalities (e.g., Computed Tomography) is also an interesting direction.